\documentclass{article}

\newtheorem{ThMain}{Theorem}
\newtheorem{Cor}{Corollar}[ThMain]

\renewcommand{\O}{\mathcal{O}}

\newcommand{\sgn}{\hbox{\rm sgn}}

\font\Bbb = msbm10.tmf
\def\R {{\hbox {\Bbb R}}}
\def\bF {{\hbox{\bf F}}}
\def\bu {{\hbox{\bf u}}}

\title
{\bf Averaging of an autoresonance model}
\author
{\bf L.A. Kalyakin
\\
Institute of Mathematics,  Ufa Sci. Centre,  Russian Acad. of
Sci.\\  112,Chernyshevsky str., Ufa,
\\450000, Russia\\
E-mail: klen@imat.rb.ru
\thanks
{The russian version of the paper is presented in "Matemat.
Zametki" ("Mathematical Notes" in English)} }
\date{July. 17. 2001}

\begin{document}

\maketitle

\begin{abstract}
Autoresonance is a phase locking phenomenon occurring in
nonlinear oscillatory system, which is forced by oscillating
perturbation. Many physical applications of the autoresonance
are known in nonlinear physics. The essence of the phenomenon
is that the nonlinear oscillator selfadjusts to the varying
external conditions so that it remains in resonance with the
driver for a long time. This long time resonance leads to a
strong increase in the response amplitude under weak driving
perturbation. An analytic treatment of a simple mathematical
model is done here by means of asymptotic analysis using a
small driving parameter. The main result is finding threshold
for entering the autoresonance.
\end{abstract}

\section{ Introduction} A hamiltonian system of two
differential equations with a small parameter $\varepsilon$
$$u^\prime-H_v(u,v)=\varepsilon f(\tau)\cos(\varphi),\quad
v^\prime+H_u(u,v)=\varepsilon g(\tau)\cos(\varphi). \eqno (1)
$$ is considered. Here the right hand side represents a
prescribed small external force which is fast oscillating:
$\varphi=\Phi(\tau)/\varepsilon$. The $f,g,\Phi(\tau)$ are
smooth functions of the slow time $\tau=\varepsilon t$. It is
supposed that the unperturbed system (as $\varepsilon=0$) has
the stable equilibrium point $(0,0)$ of center type in general
position, which is taken as the initial point of the perturbed
solution $$u,v(t;\varepsilon )\vert _{t=0}=0. \eqno (2) $$

In this note we give an asymptotic solution of the problem
(1),(2) as $\varepsilon\to 0$, which is valid for large time
$t=\O(\varepsilon^{-1})$. The main goal is to find conditions
under which the system's energy
$H(u,v)+\varepsilon[vf-ug]\cos(\varphi)$ grows up to the
order of unity while the driver is being small: $\varepsilon
f,\varepsilon g\!=\!o(1),\ \varepsilon\!\to\! 0, $. Such type
phenomenon is referred to as autoresonance and one was
studied by different authors [1-10].

Our construction is based on general two-parametric periodic
solution of the unperturbed system: $u_0,v_0(t+t_0,E),\
\allowbreak (t_0\in \R ,\ E\in(0,e_0)\subset\R)$ which is
exists in a neighborhood of the equilibrium point $(0,0)$. We
set $E=0$ in the equilibrium solution $u_0,v_0(t,0)\equiv 0$.
The parameter $E$ is a first integral of the unperturbed
system, hence one can be identified as unperturbed energy
$E=H(u,v)$.

We consider the general case, when both the period $T=T(E)$
and the frequency $\omega =2\pi /T(E)$ depend on the energy
$E$ so that $T^\prime(E)\neq 0$. This dependence is a
speciality of nonlinear system. It provides entering in
autoresonance under different driver frequency.

\section
{Anzatz} An averaging method elaborated by Kuzmak [11] is here
applied to asymptotic analysis of the solution. Sometimes this
asymptotic approach is referred to as adiabatic
approximation. The basic idea is that the leading order term
of the asymptotic solution is taken as unperturbed solution
$$\left(\begin{array}{c}u\\
v\end{array}\right)=\left(\begin{array}{c}u_0(\sigma ,E)\\
v_0(\sigma ,E)\end{array}\right)+\varepsilon
\left(\begin{array}{c}u_1(\sigma ,\tau)\\
v_1(\sigma ,\tau)\end{array}\right)+\O(\varepsilon^2),\quad
(\tau=\varepsilon t) \eqno (3)$$ under appropriate slow
deformation of both the phase $\sigma=\varepsilon
^{-1}\Psi(\tau ;\varepsilon )/\omega(E)$ and energy
$E=E(\tau,\varepsilon)$. In the leading order term the problem
is to find two functions $\Psi,E(\tau ;\varepsilon )$
depending on slow time $\tau=\varepsilon t$. The main result
is asymptotics of these functions. $$\Psi(\tau ;\varepsilon )
=\Psi_{0}(\tau )+\varepsilon \Psi_{1}(\tau
)+\varepsilon^2\Psi_{2}(\tau )+\O(\varepsilon^3), \quad
E(\tau,\varepsilon)=E_{0}(\tau )+\varepsilon E_{1}(\tau
)+\O(\varepsilon^2) \eqno (4)$$ The initial energy is zero
$E(0,\varepsilon)=0$. If we find that $E_0(\tau)=\O(1)$ as
$\tau=\O(1)$, it means that the energy of the system goes to
order of unity for large times $t=\O(\varepsilon^{-1})$.

\section{The slow deformation equation}The property of $T$ --
periodic with respect to fast variable $\sigma$ identifies a
class of functions $u,v(\sigma,\tau)$ which are used for
asymptotic solution. This secular condition, as applied to the
first order correction, leads to the equations for the
desired slow varying functions.  In the leading order term we
obtain $$\Psi^\prime_0=\omega(E_0), \quad E^\prime_0={1\over
T}\int_0^{T}[\partial_\sigma\bu_0(\sigma,E_0),\bF
(\varphi;\tau )]\, d\sigma \eqno (5)$$ Here the integrand is
skew-symmetric product of the
$\partial_\sigma\bu_0=(u_0^\prime(\sigma,E),v_0^\prime(\sigma,E))$
by the perturbation, which is $\bF=(f,g)\cos(\varphi)$ in the
case of (1) . As usually in this way the relations (5) are
looking as differential equations. But in the case under
consideration we can convert theirs into algebraic equations.
To this end one have to take into account that there is the
fast variable $\varphi=\Phi(\tau)/\varepsilon$ with
prescribed function $\Phi(\tau)$ in the original problem (1).
It is not to hard to guess that in resonance case the desired
fast variable $\sigma$ must to be related with the given
$\varphi$. Such type relation we refer to as

{\bf Resonance requirement.} {\it Difference between driver
frequency and free frequency is small}
$$\Psi^\prime-\Phi^\prime=\O(\varepsilon)\eqno (6)$$ So we try
to find asymptotic solution in the form (3),(4) under
additional relation (6). This approach may be realized only
under some restrictions which are just autoresonance
conditions.

As the leading order term of the phase $\Psi_0(\tau)$ is the
given function $\Phi_0(\tau)$ hence the first equation in (5)
is reading as an algebraic equation
$$\omega(E_0)=\Phi^\prime(\tau)\eqno (7)$$ for the energy
$E_0(\tau)$. Next, the second equation in (5) is reading as
an algebraic equation for the first order correction of the
phase function $\Psi_1(\tau)$. In order to see it we have to
take into account the resonance relation (6) as follows
$\varphi=\omega(E)\sigma-(\Psi-\Phi)/\varepsilon=
\omega(E)\sigma-\Psi_1+\O(\varepsilon)$.

Analysis of higher order corrections $(u_n,v_n),\ n\geq 2$
leads to linear algebraic equations for the corrections of
both the energy and the phase function in the expansion (4).

\section{Results} It is clear that in general case the
nonlinear algebraic equations have no solutions. Requirements,
under which they are solved, yield conditions of entering the
autoresonance.

{\ThMain{Let both $\omega^\prime(E)\neq 0$ and the driver
frequency coincide with the free frequency in the initial
moment $\Phi^\prime(0)=\omega(0)$. Then the equation (7) has
unique growing solution $E_0=E_0(\tau)$ for all
$\tau\in[0,\tau_0]$ if and only if directions of the
frequency variations coincide
$\sgn\,\Phi^{\prime\prime}(\tau)=\sgn\,\omega^\prime(E)$.}}

Now the second equation in (5) is reduced to the algebraic
equation  for the $\Psi_1$ $$-{{\Phi^{\prime}(\tau)}\over
2\pi}\int_0^{2\pi}[\bu_0(T(E_0)s/2\pi,E_0)),\partial_s\bF
(s-\Psi_1;\tau )]\, ds
=\Phi^{\prime\prime}(\tau)/\omega^\prime(E_0)  \eqno (8)$$ The
equation can be solved only under specific conditions on the
driver data $\bF(s,\tau)$, $\Phi(\tau)$. In particular, one
can see from (8) that the property:
$\Phi^{\prime\prime}(0)=0$ is need, because the unperturbed
solution tends to zero: $\bu_0(T(E_0)s/2\pi,E_0)\to 0$  as
energy tends to zero $E_0(\tau)\to 0, \ \tau\to 0$. A crucial
condition is imposed on the amplitude. In order to clear
situation we consider a specific driver as  given in (1). In
that case the (8) is reduced to the trigonometric equation as
follows $$\Phi^{\prime}(\tau)[a(\tau)\cos(\Psi_1)-
b(\tau)\sin(\Psi_1)]={{\Phi^{\prime\prime}(\tau)}
/{\omega^\prime(E_0(\tau))}} \eqno (9)$$ Here $a,b$ are
Fourier coefficients depending on the unperturbed solution
$$a={{1}\over 2\pi}\int_0^{2\pi}[g(\tau)
u_0-f(\tau)v_0]\sin(s)\,ds, \quad b={{1}\over
2\pi}\int_0^{2\pi}[g(\tau)u_0-f(\tau)v_0]\cos(s)\, ds $$ Thus
the $a,b$ are known as far as the unperturbed solution is
known: $(u_0,v_0)(\sigma,E)$ under $\sigma=T(E)s/2\pi,$
$E=E_0(\tau)$. The following assertion is easily proved.

{\ThMain{Let both the the right hand side in the original
problem be given as in (1) and conditions of theorem 1 hold.
Then equation (8) is solved if and only if the inequality
$${a^2(\tau)+b^2(\tau)}\geq
[{{\Phi^{\prime\prime}(\tau)}/{\omega\omega^\prime(E_0(\tau))}}]^2
\eqno (10)$$ holds.}}

Condition (10) may be simplified. To this end an asymptotics
of the unperturbed solution as $E\to 0$ can be used. In this
way the following assertion is obtained:

{\Cor{Let both $\Phi^{\prime\prime}(0)=0$ and the strong
inequality $$\omega^2(0)[{f^2(0)+g^2(0)}]>
4|{\Phi^{\prime\prime\prime}(0)/\omega^\prime(0)}|$$ hold in
the initial moment. Then the equation (9) has a solution
$\Psi_1=\Psi_1(\tau)$ on some finite interval
$\tau\in[0,\tau_0]$.}}

\section {Conclusion} Requirements of the theorems 1,2 may be
considered as conditions under which the system enters
autoresonance, so that the energy of the system grows up to
order of unity for large times
$t=\tau/\varepsilon=\O(\varepsilon^{-1})$. In particular
inequality (10) may be interpreted as threshold for entering
the autoresonance.

\section {Restrictions} For the first we have to remark that the
formulas (3),(4) give only a formal asymptotic solution. The
problem of justification of the asymptotics remains open.
Moreover there remains an ambiguity in the the phase shift of
the leading order term because of two roots of the equation
(9).

For the second we have to note that the formal asymptotic
solution as given above is valid only for large times and one
is not suitable on initial stage. In order to see this
nuisance one have to analyze higher order terms in the
expansion (4). They are obtained from linear algebraic
equations and in this way any additional requirements on the
data do not arise. However higher order corrections have
singularities as $\tau\to 0$, which became stronger as the
number of the correction increases. For example,
$E_1(\tau)=\O(\tau^{-1}),\ \Psi_2(\tau)=\O(\tau^{-3})$. Such
type singularities indicate that the anzatz (3),(4) is not
suitable on the initial stage, when $\tau=\varepsilon t$ is
small. Formulas (3),(4) give an asymptotic solution only for
large times $\varepsilon^{-2/3}\ll t\leq
\O(\varepsilon^{-1})$.

Asymptotic solution on the initial stage must be taken in
another form in which the slow time
$\theta=\varepsilon^{2/3}t$ is  used [12]. Such structure of
the asymptotics is similar to boundary layer. Solution which
is valid everywhere can be obtained by matching method [13].

\section {Notes to bibliography} Applications of the
phenomenon similar to autoresonance was firstly suggested by
Veksler [1,2], Andronov -- Gorelik [3] and McMillan [4]. In
particular, threshold for entering the autoresonance was
pointed in [2]. Later, different systems were studied and
entering the autoresonance was found in different cases, see
for instance [5-10]. However accurate mathematical analysis of
the autoresonance phenomenon does not undertake up to now.


\end{document}